\documentclass[twocolumn,showpacs,preprintnumbers,amsmath,amssymb]{revtex4}
\usepackage{amsmath,amssymb,graphics,epsfig,subfigure}
\usepackage{color}

\begin{document}

\thispagestyle{empty}

\begin{center}

\title{Intrinsic curvature and topology of shadows in Kerr spacetime}

\date{\today}
\author{Shao-Wen Wei$^{1,2}$ \footnote{E-mail: weishw@lzu.edu.cn} ,
        Yu-Xiao Liu$^{1}$ \footnote{E-mail: liuyx@lzu.edu.cn} , and Robert B. Mann$^{2}$ \footnote{E-mail: rbmann@uwaterloo.ca}}

 \affiliation{ $^{1}$Institute of Theoretical Physics $\&$ Research Center of Gravitation, Lanzhou University, Lanzhou 730000, People's Republic of China\\
$^{2}$Department of Physics and Astronomy, University of Waterloo, Waterloo, Ontario, Canada, N2L 3G1}

\begin{abstract}
From the viewpoint of differential geometry and topology, we investigate the characterization of the shadows in a Kerr spacetime. Two new quantities, the length of the shadow boundary and the local curvature radius are introduced. Each shadow can be uniquely determined by these two quantities. For the black hole case, the result shows that we can constrain the black hole spin and the angular coordinate of the observer by only measuring the maximum and minimum of the curvature radius. While for the naked singularity case, we adopt the length parameter and the maximum of the curvature radius. This technique is completely independent of the coordinate system and the location of the shadow, and is expected to uniquely determine the parameters of the spacetime. Moreover, we propose a topological covariant quantity to measure and distinguish different topological structures of the shadows.
\end{abstract}


\maketitle
\end{center}

{\emph{Introduction}.}---Black holes are amongst the most fascinating objects in the Universe. Recently, for the first time, the twin Laser Interferometer Gravitational-wave Observatory (LIGO) detectors directly measured gravitational waves~\cite{Abbott} due to black hole collisions, arousing great interest in gravitational waves and providing the most direct evidence for the existence of both black holes and  binary black hole systems. From this data the masses and spins of these colliding black holes can be determined.  Furthermore, it is generally believed that there exists a super-massive black hole at the center of each galaxy.  Such objects will cast a two-dimensional shadow
from light emitted by objects behind it~\cite{Falcke}, and whose particular form may reveal not only black hole
properties but perhaps even new tests of gravitational theory.   Very recently, the Event Horizon Telescope has been able to observe the shadow cast by the super-massive black hole in the center of our galaxy~\cite{Fish} at very high angular resolution, and its shape is expected to be determined with the new imaging techniques.

From a theoretical point of view, the boundary of the shadow corresponds to the photon capture sphere as seen by a distant observer. Conversely a computation of this latter structure via null geodesics determines the  black hole shadow. In order to fit  astronomical observations, several observables were constructed using special points on the shadow boundary in the celestial coordinates~\cite{Hioki,Johannsen}. Various related investigations have been carried out by many groups (e.g.~\cite{Stuchlik,Amarilla,Yazadjiev,Bambi,Wei,Atamurotov,Mann,Cunha,Younsi,Ghasemi,Lammerzahl,Paganini}). Although the velocity of the observer can yield aberrational effects, these were found to be negligible for observing the shadow of Sgr A*~\cite{Grenzebach}.  A formalism for describing the boundary of the shadow,  independent of the location its center, has been proposed~\cite{Abdujabbarov}, and several distortion parameters introduced. This method is expected to give a possible precise detection of the black hole shadow.

How to precisely describe the boundary of the black hole shadow is therefore crucial for measuring black hole parameters with astronomical observations. It is known that the boundary of the shadow cast by a black hole is a one-dimensional closed curve. However, other dark objects, in particular naked singularities,  have
shadow boundaries that are open curves. Fortunately differential geometry provides a natural way to precisely describe the intrinsic properties of  closed and open curves (i.e. surfaces and curves).  Here we employ such methods to study and compare
these two kinds of shadows, introducing  covariant quantities to measure the topological structure of a shadow.
Our approach can therefore not only distinguish different black hole and naked singularity backgrounds but can also reveal any potential topological phase transition between these systems.

\begin{figure}
\includegraphics[width=7cm]{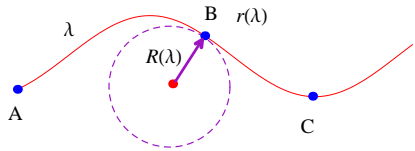}
\caption{Sketch picture. The curve starting at point A is parametrized by the line length parameter $\lambda$ and local curvature radius $R(\lambda)$. The circle denotes the curvature circle with radius $R(\lambda)$ at point B.}\label{skip}
\end{figure}

{\emph{Set up}.}---The pattern of a shadow is characterized by its boundary, a one-dimensional curve. By determining the nature of the curve, we will obtain the basic properties of a black hole or naked singularity. From the viewpoint of differential geometry, a curve has two intrinsic quantities: its length $\lambda_{\text{s}}$ and its local curvature radius $R$. A sketch is provided in Fig.~\ref{skip}, where $\lambda$  parametrizes the curve.
 Using $(\lambda_{\text{s}} , R)$, a curve will be uniquely determined  without introducing any coordinate system. Note that if the curvature radius at point B is defined to be positive, then the one at point C will be negative.

\begin{figure}
\includegraphics[width=6cm]{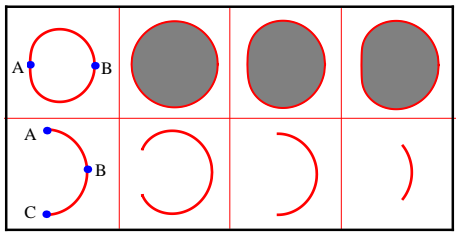}
\caption{Shadows for the Kerr black hole (upper row) and Kerr naked singularity (lower row). The spin $a$ increases from left to right.}\label{skip2}
\end{figure}
\begin{figure}
\includegraphics[width=6cm]{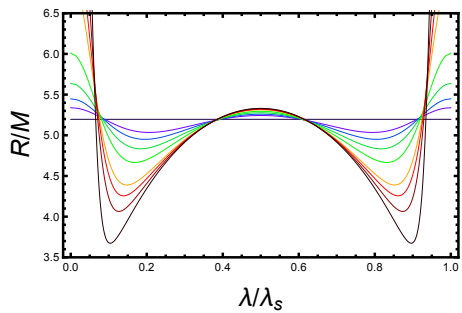}
\caption{The local curvature radius $R$ as a function of the line length parameter $\lambda$. Near the well, the spin $a/M$=0, 0.01, 0.5, 0.6, 0.7, 0.8, 0.9, 0.93, 0.96, 0.99 from top to bottom, with the horizontal black line
being the Schwarzschild case.}\label{skip3}
\end{figure}

{\emph{Black hole shadow}.}---Here we consider a simple case in which all light sources are located at infinity and  distributed uniformly in all directions. The observer is likewise located at infinity. The rotating dark object has inclination angle $\theta_{0}$,   defined by the angle between its rotation axis and the observer's line of sight. Then the celestial coordinates describing its shadow are~\cite{Bardeen,Chandrasekhar}
\begin{eqnarray}
 \alpha&=&-\xi \csc\theta_{0}, \\
 \beta&=&\pm\sqrt{\eta+a^{2}\cos^{2}\theta_{0}-\xi^{2}\cot^{2}\theta_{0}},
\end{eqnarray}
where $a$ is the black hole spin, and $\xi = L_z/E$ and $\eta = \mathcal{Q}/E^{2}$ are two conserved parameters for the photon orbit associated respectively with the angular momentum and Carter constant of the null geodesic of energy $E$. On the boundary of the shadow, the parameters $\xi$ and $\eta$ are parametrized by the radius of the circular unstable photon orbit $r_{0}$~\cite{Bardeen,Chandrasekhar}
\begin{eqnarray}
 \xi&=&\frac{(3M-r_{0})r_{0}^{2}-a^{2}(M+r_{0})}{a(r_{0}-M)},\label{xx}\\
 \eta&=&\frac{r_{0}^{3}(4a^{2}M-r_{0}(3M-r_{0})^{2})}{a^{2}(r_{0}-M)^{2}}.\label{ee}
\end{eqnarray}

We show the shapes of the shadow in this coordinate for Kerr black holes (upper row) and Kerr naked singularities (lower row) in Fig.~\ref{skip2}. For the black hole the shape is a two-dimensional dark region (whose boundary is topologically the same as a closed circle), whereas it is a one-dimensional dark arc for the singularity. The boundary of each shadow has a $\mathcal{Z}_{2}$ symmetry, which gives
\begin{eqnarray}
 R(\lambda_{\text{s}}-\lambda)=R(\lambda).
\end{eqnarray}
Moreover, for a black hole shadow, one has
\begin{eqnarray}
 R(\lambda+\lambda_{\text{s}})=R(\lambda),
\end{eqnarray}
with $\lambda_{\text{s}}$ being the length of the boundary of the shadow.

Consider an observer located at the equatorial plane $\theta_{0}=\pi/2$ of the black hole spacetime background.
On this plane, $r_{0}\in(r_{\text{A}},\;r_{\text{B}})$ with
\begin{eqnarray}
 r_{\text{A, B}}=2M\left(1+\cos\left(\frac{2}{3}\arccos\left(\mp\frac{a}{M}\right)\right)\right),
 \label{rr}
\end{eqnarray}
respectively corresponding to direct and retrograde orbits. The celestial coordinates  reduce to $\alpha=-\xi(r_{0})$, $\beta=\pm\sqrt{\eta(r_{0})}$.
For the Schwarzschild black hole, $a=0$ yielding $r_{0}=3M$ and $\alpha^{2}+\beta^{2}=(3\sqrt{3}M)^{2}$, which indicates that the shadow of the Schwarzschild black hole is a standard circle with radius $3\sqrt{3}M$. For $a\neq0$, taking this parametrization, we can calculate the length $\lambda_{\text{s}}$ of the shadow boundary with
\begin{eqnarray}
 \lambda_{s}=2\int_{r_{\text{A}}}^{r_{\text{B}}}
   \sqrt{(\partial_{r_{0}}\alpha)^{2}+(\partial_{r_{0}}\beta)^{2}}dr_{0}.
\end{eqnarray}
The factor `2' comes from the $\mathcal{Z}_{2}$ symmetry. A simple calculation shows that $\lambda_{\text{s}}$ decreases with the black hole spin $a$. However, the difference is very small.  For example, when $a/M=0$, 0.99, and 1, we have $\lambda_{\text{s}}/M$=$6\pi\sqrt{3}$, 31.2636, and $16\sqrt{3}$, respectively. The relative deviation between $a/M=0$ and 0.99 is about 4.3\%.

Alternatively, the local curvature radius at each point can also be parametrized by $r_{0}$.  Taking nearby points
$r_{0}\pm\epsilon$, these three points on the boundary of shadow can be used to uniquely construct a circle. As $\epsilon\rightarrow0$, the radius of the circle will exactly coincide with the local curvature radius at the point corresponding to $r_{0}$. Fortunately it has an analytic form
\begin{eqnarray}\label{R0}
 R=\frac{8\sqrt{Mr_{0}}(r_{0}^{3}-3Mr_{0}^{2}+3M^{2}r_{0}-Ma^{2})}{3(r_{0}-M)^{3}}.
\end{eqnarray}

We plot the curvature radius $R$ as a function of the length parameter $\lambda$ in Fig.~\ref{skip3} for different values of the black hole spin $a$ when point A (shown in Fig.~\ref{skip2}) is chosen as the starting point. Since our treatment is independent of the coordinate system, the starting point can be chosen arbitrarily,  with the pattern of the curvature radius $R$ unchanged. For $a/M=0$, we can see that the curvature radius $R=3\sqrt{3}M$, which means that the black hole shadow is just a standard circle. For nonvanishing spin $a/M\neq0$, the curvature radius has two maxima at $\lambda$=0 and $\lambda_{\text{s}}$ and two minima at some $\lambda$ satisfying $\partial_{r_{0}}R=0$, due to the $\mathcal{Z}_{2}$ symmetry. Reflecting about the symmetric point, the curvature radius will have only one maximum and one minimum for each fixed spin $a$. Moreover, the minimum value of $R$ decreases and shifts to smaller $\lambda$ with increasing spin,  approaching its (approximate) minimum value 2.7$M$ for an extremal black hole with $a/M=1$. Conversely the maximum
value of $R$ increases with the black hole spin and diverges for the extremal black hole. In general each black hole shadow is characterized by  maximal and minimal values of $R$, providing a way to measure  black hole spin  whilst ignoring the length of the shadow boundary.

\begin{figure}
\includegraphics[width=6cm]{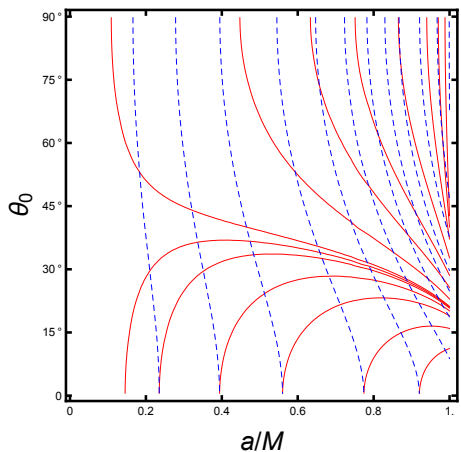}
\caption{Contour lines of $R_{\text{min}}$ and $R_{\text{max}}$ in $\theta_{0}$-$a$ plane for a Kerr black hole shadow. $R_{\text{max}}/M$ is described by the red solid lines with fixed values 4.9, 5, 5.1, 5.15, 5.18, 5.19, 5.2, 5.3, 5.5, 5.8, 6.5, 8, 10, 14 from bottom to top, whereas $R_{\text{min}}/M$ is described by the blue dashed lines with fixed values 3.2, 4.0, 4.3, 4.5, 4.6, 4.7, 4.8, 4.9, 5.0, 5.1, 5.15, 5.18 from right to left.}\label{skip4}
\end{figure}

For the observer located at $\theta_{0}\neq\pi/2$ the situation is similar. The celestial coordinates can also be parametrized by the radius of the circular unstable photon orbit, i.e.,  $\alpha=\alpha(r_{0})$ and $\beta=\beta(r_{0})$, but $r_{0}$ is no longer in the range $(r_{\text{A}},\;r_{\text{B}})$ given in Eq.~(\ref{rr});  instead it is determined by solving $\beta(r_{\text{A,B}})=0$. The local curvature radius can also be expressed in terms of $r_{0}$
\begin{widetext}
\begin{eqnarray}
 R=\frac{64M^{1/2}(r_{0}^{3}-a^{2}r_{0}\cos^{2}\theta_{0})^{3/2}\left[r_{0}(r_{0}^{2}-3Mr_{0}+3M^{2})-a^{2}M^{2}\right]}
   {(r_{0}-M)^{3}\left[3(8r_{0}^{4}-a^{4}-8a^{2}r_{0}^{2})-4a^{2}(6r_{0}^{2}+a^{2})\cos(2\theta_{0})
    -a^{4}\cos(4\theta_{0})\right]},
\label{Rad00}
\end{eqnarray}
\end{widetext}
which reduces to \eqref{R0} for $\theta_{0}=\pi/2$.
From the above expression, we can obtain the maximum $R_{\text{max}}$ and minimum $R_{\text{min}}$ of the curvature radius. In Fig.~\ref{skip4}, we show the contour lines of $R_{\text{max}}$ (red solid lines) and $R_{\text{min}}$ (blue dashed lines). For fixed viewing angle $\theta_{0}$, we  see that the black hole spin can be uniquely determined by only measuring $R_{\text{min}}$. However, it cannot be uniquely determined by $R_{\text{max}}$ if $\theta_{0}$ is approximately in the range of $(17^{\circ}, 40^{\circ})$. Nevertheless, each point in the figure is characterized by a pair of values $(R_{\text{min}},\;R_{\text{max}})$. So we can easily fix $\theta_{0}$ and $a$ by measuring $R_{\text{min}}$ and $R_{\text{max}}$ for the black hole.

{\emph{Naked singularity shadow}.}---When the spin of the Kerr black hole is beyond its extremal value, i.e.,
\begin{eqnarray}
 |a|>M,
\end{eqnarray}
the system becomes a naked singularity.  Although naked singularities have generally been thought to be 
unstable~\cite{Gleiser}  -- indeed,  the cosmic censorship conjecture forbids them~\cite{Penrose} --
recent work has shown that when  scalar fields or other fields are included, stability can be 
restored~\cite{Sadhu}, and the cosmic censorship conjecture will be violated~\cite{Destounis}. 
In particular, it has recently been claimed (using an analytical treatment employing the Teukolsky equation)  that Kerr naked singularities are stable under a variety of boundary conditions~\cite{Nakao}. Thus it is of great interest to study the naked singularity shadow.

As mentioned above, there exists a dramatic change of the shadow shape and topology for the naked singularity. The boundary closed curve will become open, and will be a one-dimensional dark arc. In realistic observations it inclines to form a two-dimensional dark ``lunate'' shadow; observation of such shapes would provide evidence for the existence of naked singularities  in our Universe.

\begin{figure}
\includegraphics[width=6cm]{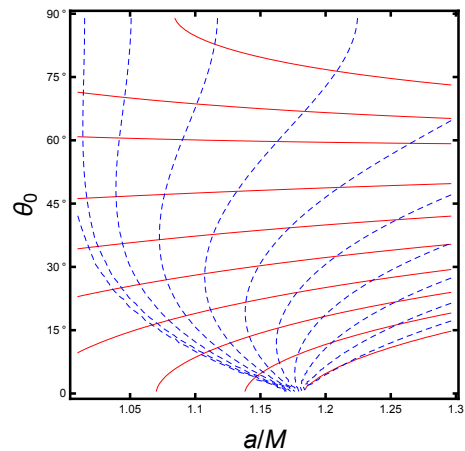}
\caption{Contour lines of $R_{\text{max}}$ and $\lambda_{\text{s}}$ in $\theta_{0}$-$a$ plane for a Kerr naked singularity. $R_{\text{max}}/M$ is described by the red solid lines with fixed values 4.55, 4.65, 4.75, 4.85, 4.95, 5.05, 5.15, 5.25, 5.3, 5.35 from bottom to top, whereas $\lambda_{\text{s}}/M$ is described by the blue dashed lines with fixed values 10.8, 12.4, 14, 15.6, 17.2, 18.8, 20.4, 22, 23.6, 25.2, 26.2 from right to left.}\label{skip5}
\end{figure}

For this case, we can also calculate  the curvature radius $R$ and boundary length $\lambda_{\text{s}}$. After a detailed examination, we find that the curvature radius $R$ increases with the length parameter $\lambda$. So for each naked singularity shadow, $R$ has one minimum at point A or C and one maximum at point B (shown in Fig.~\ref{skip2}). Thus, similar to the black hole shadow, we can measure  $a$ and $\theta_{0}$ by observing the minimum and maximum of the curvature radius. However, the minimum curvature radius corresponding to the starting point A or C is very hard to calculate. We therefore make use of the length $\lambda_{\text{s}}$ of the shadow, as well as the maximum of $R$ to determine the shadow. In Fig.~\ref{skip5} we list the contours of $R_{\text{max}}$ and $\lambda_{\text{s}}$.  Interestingly, if $\theta_{0}$ is fixed, we can determine the spin $a$ of the naked singularity, which is very different from the black hole case. Note that a measurement only of $\lambda_{\text{s}}$ cannot do that. One can accurately obtain the spin $a$ and inclination angle $\theta_{0}$ by measuring both $R_{\text{max}}$ and $\lambda_{\text{s}}$.

{\emph{Topology}.}---While the black hole and naked singularity have distinct shadows that can be distinguished   by observation,  we are also interested in seeking some topological covariant quantities to describe their difference.  The boundary of each shadow is in a one-dimensional curve, and  cannot cross  itself. Motivated by the Gauss-Bonnet-Chern theorem we introduce the topological covariant quantity
\begin{eqnarray}
 \delta=\frac{1}{2\pi}\bigg(\int \frac{d\lambda}{R(\lambda)}+\sum_{i}\theta_{i}\bigg),
 \label{delta}
\end{eqnarray}
 where $\lambda$ and $R(\lambda)$ are the line length parameter and the local curvature radius we introduced above. The first term measures the smooth part of the shadow boundary and the parameter $\theta_{i}$ denotes the $i$-th angle of the boundary. Since the shadow boundary of the Kerr black hole is smooth,  $\sum_{i}\theta_{i}$=0. Cases for which  $\theta_{i}$   is non-vanishing are related to the existence of   stable fundamental photon orbits; examples include   scalar hair black holes~\cite{Cunha}, rotating Konoplya-Zhidenko black holes, or a compact object with magnetic dipole~\cite{Wang,Jing}. However, for the Kerr spacetime, all the fundamental photon orbits are unstable, and so the second term in Eq.~(\ref{delta}) vanishes.

\begin{figure}
\includegraphics[width=6cm]{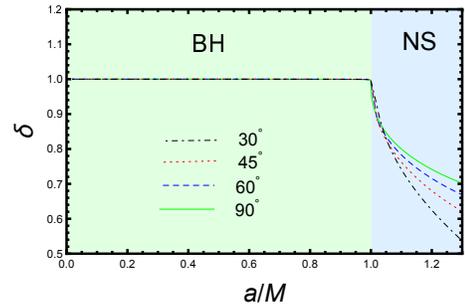}
\caption{Behavior of the topological covariant quantity $\delta$ as a function of spin $a$. The viewing angle is set as $\theta_{0}$=$30^{\circ}$, $45^{\circ}$, $60^{\circ}$, $90^{\circ}$, respectively. Light green (left) region denotes the black hole region of $a\leq M$, and light blue (right) region denotes the naked singularity region of $a>M$.}\label{skip6}
\end{figure}

We exhibit in  Fig.~\ref{skip6}
the topological quantity $\delta$ for various fixed $\theta_{0}$=$30^{\circ}$, $45^{\circ}$, $60^{\circ}$, $90^{\circ}$. For the black hole ($a\leq M$), $\delta$ keeps a constant value 1, whereas for the naked singularity ($a>M$), $\delta$ deviates from unity, rapidly decreasing with increasing spin $a$. For $a>M$ there is a  naked singularity, and the shadow experiences a structural change from a two-dimensional dark region to a one-dimensional dark lunate. As the dimension of the structure changes, this is a topological change and so the quantity $\delta$ can be regarded as a topological covariant quantity whose value is indicative of  such topological change. Noting that $R(\lambda)$ might be negative (at point C in Fig.~\ref{skip}), $\delta$ can also be used to measure the structural change of the shadow of a scalar hair black hole~\cite{Cunha}, as well as for more `square' or `hammer-like' shadows.  For multiple disconnected shadows~\cite{Mann,Yumoto}, our topological quantity  gives
\begin{eqnarray}
 \delta= n,
\end{eqnarray}
where $n$ is the number of  disconnected shadows. So different values of $\delta$ are indicative of different topological structures of the shadows. Note that in Ref.~\cite{Oancea}, the authors have discussed the the topological structure of the past and future trapped null geodesics, however which is quite different from our case.


{\emph{Summary}.}---Making use of concepts from  differential geometry we have constructed a novel approach for determining the shadow cast by a rotating black hole and its naked singularity counterpart. Our approach depends only on intrinsic properties of the shadow, and is independent of the coordinate system and the location of the shadow.

We found that both the spin $a$ and viewing angle $\theta_{0}$ can be accurately determined by measuring the maximum and minimum of the local curvature radius $R(\lambda)$ of the shadow boundary of the black hole. The minimal value of $R$ is difficult to obtain for the naked singularity, but this can be dealt with by measuring the length of the lunate shadow and the maximum of $R$ to determine $a$ and $\theta_{0}$. A full description of the shadow entails measuring the local curvature radius at each point of its boundary.

We have also introduced a topological covariant quantity, $\delta$, that characterizes the topological structure of the shadow. This quantity distinguishes between naked singularities and black holes, and is sensitive to shadow connectivity.   We expect that  our technique will have considerable advantage in fitting theoretical models to the astronomical observations.

 Of course all astronomical observations have  finite resolution, which impedes the ability to accurately measure 
differential properties of a shadow boundary, particularly the last term in \eqref{delta}.   While it is worth
investigating how to use   integral characteristics (while controlling the relative error) to determine the shadow,
we note that  the latter term in \eqref{delta} is zero
for the class of Kerr black holes, and so any evidence of a nonzero value provides significant information about
the black hole that generates it.  This can provide us with an additional tool to precisely distinguish GR from alternative gravity theories. Moreover, it is conceivable that improved resolution could better distinguish or rule out this feature, particularly when lensing rings around the shadow are taken into account. These  are easily observed due to their high brightness, and the inner one is extremely close to the edge of the shadow. It is therefore natural to identify the inner lensing ring as the boundary of the shadow; in combination with the curvature radius, one may determine the corresponding black hole parameters.

{\emph{Acknowledgements}.}---This work was supported by the National Natural Science Foundation of China (Grants No. 11675064, No. 11875175, and No. 11522541) and the Natural Sciences and Engineering Research Council of Canada. S.-W. Wei was also supported by the Chinese Scholarship Council (CSC) Scholarship (201806185016) to visit the University of Waterloo.


\begin{thebibliography}{99}
\bibitem{Abbott}
 B. P. Abbott \emph{et al} (Virgo, LIGO Scientific),
 {\em Observation of Gravitational Waves from a Binary Black Hole Merger},
   Phys. Rev. Lett. \textbf{116}, 061102 (2016), [arXiv:1602.03837[gr-qc]];
 B. P.  Abbott \emph{et al} (Virgo,  LIGO  Scientific),
 {\em GW151226: Observation of Gravitational Waves from a 22-Solar-Mass Binary Black Hole Coalescence},
 Phys.  Rev. Lett. \textbf{116}, 241103 (2016), [arXiv:1606.04855 [gr-qc]];
 B. Abbott \emph{et al} (VIRGO, LIGO Scientific),
  {\em GW170104: Observation of a 50-Solar-Mass Binary Black Hole Coalescence at Redshift 0.2},
  Phys. Rev. Lett. \textbf{118}, 221101 (2017), [arXiv:1706.01812[gr-qc]];
 B. Abbott \emph{et al}, (VIRGO, LIGO Scientific),
  {\em GW170814: A three-detector observation of gravitational waves from a binary black hole coalescence},
  Phys. Rev. Lett. \textbf{119}, 141101 (2017), [arXiv:1709.09660[gr-qc]];
 B. Abbott \emph{et al}, (VIRGO, LIGO Scientific),
  {\em GW170817: Observation of Gravitational Waves from a Binary Neutron Star Inspiral},
  Phys. Rev. Lett. \textbf{119} 161101 (2017), [arXiv:1710.05832[gr-qc]].

\bibitem{Falcke}
 H. Falcke, F. Melia, and E. Agol,
  {\em Viewing the Shadow of the Black Hole at the Galactic Center},
  Astrophys. J., \textbf{528}, L13 (2000), [arXiv:astro-ph/9912263].

\bibitem{Fish}
 V. L. Fish \emph{et al},
  {\em Observing-and Imaging-Active Galactic Nuclei with the Event Horizon Telescope},
     Galaxies \textbf{4}, 54 (2016), [arXiv:1607.03034[astro-ph.IM]].

\bibitem{Hioki}
 K. Hioki and K.-i. Maeda,
  {\em Measurement of the Kerr Spin Parameter by Observation of a Compact Object's Shadow},
   Phys. Rev. D \textbf{80}, 024042 (2009), [arXiv:0904.3575[astro-ph.HE]].

\bibitem{Johannsen}
 T. Johannsen,
  {\em Photon Rings around Kerr and Kerr-like Black Holes},
    	Astrophys. J. \textbf{777}, 170 (2013), [arXiv:1501.02814[astro-ph.HE]].

\bibitem{Stuchlik}
 J. Schee and Z. Stuchlik,
  {\em Optical phenomena in brany Kerr spacetimes},
   Int. J. Mod. Phys. D \textbf{18}, 983 (2009), [arXiv:0810.4445 [astro-ph]];
 Z. Stuchlik and J. Schee,
  {\em Appearance of Keplerian discs orbiting Kerr superspinars},
   Class. Quant. Grav. \textbf{27}, 215017 (2010), [arXiv:1101.3569 [gr-qc]].

\bibitem{Amarilla}
 L. Amarilla and E. F. Eiroa,
  {\em Shadow of a rotating braneworld black hole},
   Phys. Rev. D \textbf{85}, 064019 (2012), [arXiv:1112.6349[gr-qc]];
 L. Amarilla and E. F. Eiroa,
  {\em Shadow of a Kaluza-Klein rotating dilaton black hole},
   Phys. Rev. D \textbf{87}, 044057 (2013), [arXiv:1301.0532[gr-qc]].

\bibitem{Yazadjiev}
 P. G. Nedkova, V. K. Tinchev, and S. S. Yazadjiev,
  {\em The Shadow of a Rotating Traversable Wormhole},
    Phys. Rev. D \textbf{88}, 124019 (2013), [arXiv:1307.7647[gr-qc]];
 T. Vetsov, G. Gyulchev, and S. Yazadjiev,
 {\em Shadows of Black Holes in Vector-Tensor Galileons Modified Gravity},
 [arXiv:1801.04592 [gr-qc]].

\bibitem{Bambi}
 C. Bambi and N. Yoshida,
  {\em Shape and position of the shadow in the $\delta$=2 Tomimatsu-Sato space-time},
   Class. Quant. Grav. \textbf{27}, 205006 (2010), [arXiv:1004.3149[gr-qc]];
   N. Tsukamoto, Z. Li, and C. Bambi,
   {\em Constraining the spin and the deformation parameters from the black hole shadow},
   JCAP \textbf{1406}, 043 (2014), [arXiv:1403.0371[gr-qc]].

\bibitem{Wei}
 S.-W. Wei and Y.-X. Liu,
  {\em Observing the shadow of Einstein-Maxwell-Dilaton-Axion black hole},
   JCAP \textbf{11}, 063 (2013), [arXiv:1311.4251[gr-qc]];
   S.-W. Wei, P. Cheng, Y. Zhong, and X.-N. Zhou,
   {\em Shadow of noncommutative geometry inspired black hole},
   JCAP \textbf{08}, 004 (2015), [arXiv:1501.06298 [gr-qc]].

\bibitem{Atamurotov}
 F. Atamurotov, A. Abdujabbarov, and B. Ahmedov,
  {\em Shadow of rotating non-Kerr black hole},
   Phys. Rev. D \textbf{88}, 064004 (2013);
   A. Abdujabbarov, M. Amir, B. Ahmedov, and S. G. Ghosh,
   {\em Shadow of rotating regular black holes},
   Phys. Rev. D \textbf{93}, 104004 (2016), [arXiv:1604.03809[gr-qc]].

\bibitem{Mann}
 S. Abdolrahimi, R. B. Mann, and C. Tzounis,
  {\em Distorted Local Shadows},
   Phys. Rev. D \textbf{91}, 084052 (2015), [arXiv:1502.00073[gr-qc]];
   S. Abdolrahimi, R. B. Mann, and C. Tzounis,
   {\em Double Images from a Single Black Hole},
   Phys. Rev. D \textbf{92}, 124011 (2015), [arXiv:1510.03530[gr-qc]].

\bibitem{Cunha}
 P. V. P. Cunha, C. A. R. Herdeiro, E. Radu, and H. F. Runarsson,
  {\em Shadows of Kerr black holes with scalar hair},
   Phys. Rev. Lett. \textbf{115}, 211102 (2015), [arXiv:1509.00021[gr-qc]];
  P. V. P. Cunha, C. A. R. Herdeiro, and E. Radu,
   {\em Fundamental photon orbits: black hole shadows and spacetime instabilities},
   Phys. Rev. D \textbf{96}, 024039 (2017), [arXiv:1705.05461[gr-qc]];
  P. V. P. Cunha, C. A. R. Herdeiro, and M. J. Rodriguez, {\em Does the black hole shadow probe the event horizon geometry?}, 
  Phys. Rev. D \textbf{97}, 084020 (2018), [arXiv:1802.02675 [gr-qc]].

\bibitem{Younsi}
 Z. Younsi, A. Zhidenko, L. Rezzolla, R. Konoplya, and Y. Mizuno,
  {\em A new method for shadow calculations: application to parameterised axisymmetric black holes},
   Phys. Rev. D \textbf{94}, 084025 (2016), [arXiv:1607.05767[gr-qc]];
  R. A. Konoplya and Z. Stuchl\'{\i}k,
   {\em Are eikonal quasinormal modes linked to the unstable circular null geodesics?},
   Phys. Lett. B \textbf{771}, 597 (2017), [arXiv:1705.05928[gr-qc]].

\bibitem{Ghasemi}
 M. Ghasemi-Nodehi, Z. Li, and C. Bambi,
  {\em Shadows of CPR black holes and tests of the Kerr metric},
   Eur. Phys. J. C \textbf{75}, 315 (2015), [arXiv:1506.02627[gr-qc]].

\bibitem{Lammerzahl}
 A. Grenzebach, V. Perlick, and C. Lammerzahl,
  {\em Photon regions and shadows of Kerr-Newman-NUT black holes with a cosmological constant},
   Phys. Rev. D \textbf{89}, 124004 (2014), [arXiv:1403.5234 [gr-qc]].

\bibitem{Paganini}
 M. Mars, C. F. Paganini, and M. A. Oancea,
  {\em The fingerprints of black holes-shadows and their degeneracies},
   Class. Quant. Grav. \textbf{35}, 025005 (2018), [arXiv:1710.02402 [gr-qc]].

\bibitem{Grenzebach}
 A. Grenzebach,
  {\em Aberrational effects for shadows of black holes},
   Fund. Theor. Phys. \textbf{179}, 823 (2015), [arXiv:1502.02861 [gr-qc]].

\bibitem{Abdujabbarov}
 A. A. Abdujabbarov, L. Rezzolla, and B. J. Ahmedov,
  {\em A coordinate-independent characterization of a black hole shadow},
   MNRAS, \textbf{454}, 2423 (2015), [arXiv:1503.09054[gr-qc]].

\bibitem{Bardeen}
 J. M. Bardeen,
 {\em in Black holes},
 in Proceeding of the Les Houches Summer School, Session 215239, edited by C. De Witt and B.S. De Witt and B.S. De Witt (Gordon and Breach, New York, 1973).

\bibitem{Chandrasekhar}
 S. Chandrasekhar,
 {\em The mathematical theory of black holes}, (Oxford University Press, New York, 1992).
 

\bibitem{Gleiser}
 G. Dotti, R. Gleiser, and J. Pullin,
 {\em Instability of charged and rotating naked singularities},
 Phys. Lett. B \textbf{644}, 289 (2007), [arXiv:gr-qc/0607052];
 G. Dotti, R. J. Gleiser, J. Pullin, I. F. Ranea-Sandoval, and H. Vucetich,
  {\em Instabilities of naked singularities and black hole interiors in General Relativity},
    Int. J. Mod. Phys. A \textbf{24}, 1578 (2009), [arXiv:0810.0025 [gr-qc]].

\bibitem{Penrose}
 R. Penrose, {\em Gravitational collapse: The role of general relativity, Rivista del Nuovo Cimento}, Numero Speziale I, 257 (1969); {\em “Golden Oldie”: Gravitational Collapse: The Role of General Relativity}, Gen. Relativ. Grav. 34, 1141 (2002).

\bibitem{Sadhu}
 A. Sadhu and V. Suneeta,
  {\em A naked singularity stable under scalar field perturbations},
    Int. J. Mod. Phys. D \textbf{22}, 1350015 (2013), [arXiv:1208.5838 [gr-qc]].

\bibitem{Destounis}
 V. Cardoso, J. L. Costa, K. Destounis, P. Hintz, and A. Jansen,
  {\em Quasinormal modes and strong cosmic censorship},
    Phys. Rev. Lett. \textbf{120}, 031103 (2018), [arXiv:1711.10502 [gr-qc]].

\bibitem{Nakao}
 K.-i. Nakao, P. S. Joshi, J.-Q. Guo, P. Kocherlakota, H. Tagoshi, T. Harada, M. Patil, and A. Krolak,
  {\em On the stability of a superspinar},
    Phys. Lett. B \textbf{780}, 410 (2018), [arXiv:1707.07242 [gr-qc]].

\bibitem{Wang}
 M. Wang, S. Chen, and J. Jing,
  {\em Shadow casted by a Konoplya-Zhidenko rotating non-Kerr black hole},
    JCAP \textbf{10}, 051 (2017), [arXiv:1707.09451[gr-qc]].

\bibitem{Jing}
 M. Wang, S. Chen, and J. Jing,
  {\em Shadows of a compact object with magnetic dipole by chaotic lensing},
   Phys. Rev. D \textbf{97}, 064029 (2018), [arXiv:1710.07172[gr-qc]].

\bibitem{Yumoto}
 A. Yumoto, D. Nitta, T. Chiba, and N. Sugiyama,
  {\em Shadows of Multi-Black Holes: Analytic Exploration},
    Phys. Rev. D \textbf{86}, 103001 (2012), [arXiv:1208.0635[gr-qc]].

\bibitem{Oancea}
 C. F. Paganini and M. A. Oancea,
  {\em Smoothness of the future and past trapped sets in Kerr-Newman-Taub-NUT spacetimes},
    Class. Quant. Grav. \textbf{35}, 067001 (2018), [arXiv:1710.02403 [gr-qc]].
\end{thebibliography}
\end{document}